\documentclass[10pt]{article}

\usepackage{mathrsfs}
\usepackage{epic, eepic}
\usepackage{subfigure}
\usepackage{amsfonts, amssymb, amsmath, graphicx, epsfig, lscape, capt-of}
\usepackage{url}
\usepackage{ifpdf}
\usepackage[boxed]{algorithm2e}
\usepackage{array}
\usepackage{booktabs}
\usepackage{multirow}
\usepackage{xcolor}
\usepackage{adjustbox}

\def\inst#1{$^{#1}$}

\makeatletter

\begin{document}%


\title{Hagiotoponyms in France: Saint popularity,  like \\a herding phase transition}

\author{%
Marcel Ausloos\inst{1,2,3}
  }
\date{\today}%

\maketitle

\begin{center}
{\footnotesize
\inst{1}  GRAPES\\
Sart Tilman, B-4031 Angleur Liege, Belgium\\
\texttt{marcel.ausloos@ulg.ac.be}\\
\vspace{0.3cm} \inst{2} Department of Statistics and Econometrics \\
Bucharest University of Economic Studies\\
6 Piata Romana, 1st district, Bucharest, 010374 Romania\\
\texttt{marcel.ausloos@ase.ro}\\
\vspace{0.3cm} \inst{3} School of Business, University of
Leicester \\ Brookfield,  Leicester, LE2 1RQ, UK\\
\texttt{ma683@le.ac.uk}\\
}

\end{center}

\begin{abstract}
A spectacular order-order-like transition is presented in the   distribution of hagiotoponyms in France. Data analysis and displays distinguish male and female cases.  
The respective hapax values point  to a  very large variety of saints with a specific devotion. The most popular ones are  St. Martin and the apostles. The less popular ones are not so well known.
These features are explained in terms of herding in agent behaviors:  people have either preferred popular saints with supposedly good links to God, whence a herding behavior, or (non-herding) agents have preferred to name their local human settlement through a reference to some holy person(s) with more local specificities, - yet with moral or religious leadership, and conjectured to have good contact with God, whence at least locally  defined as a saint.
\end{abstract}


France; 

power law; 

Bradford-Zipf-Mandelbrot law;  

herding; 

 phase transition

\vskip5cm
 \section{Foreword}
Dietrich Stauffer  was very keen in propagating basic statistical mechanics ideas and models toward other branches of sciences \cite{DS1,DS2,DS3,stauffer1998Canpercolationtheory}.  Here is a "sociological case", an interdisciplinary research,  he might have liked,  and enjoyed it, since it concerns some data analysis of some "exotic", unusual,  data.  The analysis seems to indicate that there is something like a phase transition, because, on a log-log plot, one finds two straight lines, with different slopes on both sides of a critical point,  as are critical exponents  in research on thermodynamic and geometric {\it bona fide} phase transitions.

This type of (so called) phase transition for which I will give some tentative explanation seems to bear upon  sociological behavior of likely religious populations; at least, one can surely conjecture  some religiosity affair in the historical and geographic  context at work here.   I call the finding a "herding behavior phase transition". The words have already been used for the behavior of investors on the  financial markets \cite{Banerjee,EZmodel,guo2009herd,kononovicius2013threestates}. But there is not much relationship with such an investor phenomenon here below.

What I found, see Fig. \ref{FZipf3},  was briefly shown at some small gathering, some time ago, but  has never been submitted for publication,  

The origin of the data and analysis will be outlined in subsequent sections below. However,  preliminary remarks seem appropriate. They explain  the "why?" of this "research", as Dietrich would have surely liked. First, I should recall that I was educated in catholic schools, and interested in history and geography. Moreover, there are many festivals, e.g. in Western Europa at least,  depending on days in which (so called) saints are honored and  implored for transmitting to God some request or huge thanks from people.  Devotion to saints is an integral part of catholicism ($https://the-shrine.org/resources/catholic-devotion-to-the-saints/$). In contrast to protestant  movements, - which promote a direct connection with God,  it is  not wrong for catholics to pray  a saint for his or her intercession to God.  Saints   play a major role in the Orthodox churches as well, but the examined data does not pertain to orthodox tradition, thereafter.

Another "detail" has to be noted. Up to recent times the first name of children were often saints' names given at baptism.
 Thus  famous Saints, like St. Peter, St. Paul, St. Jean (John), St. Etienne (Stephen), St. Marcel, St. Roch, and for women, St. Marie (Mary\footnote{Officially, the "Virgin Mary", St;  Marie is not a "canonized" Sainte, because having been carried up into heaven, there is no available relics.}) , Ste. Élisabeth (Elizabeth), Ste. Catherine, Ste. C\'ecile, were quite popular "first names".
Traveling in (Southern)  France,  I came across villages  called St. Lions, St. Jurson,  St. Usuge,   St. Latier,  St. Avit, and even St. Saturnin (sometimes called San Savournin).

It was appealing to consider that  many other Saints of whom I was aware  were somewhat famous (for not necessarily religious reasons), with some singularity, like jet set, wine, cheese, ...,  as St. Tropez,  St.  Malo, St. Emilion, St. Est\`ephe, St. Véran, St. Nectaire, ... . Going to the touristic literature, I "came across" Saint-Remy-en-Bouzemont-Saint-Genest-et-Isson (F-51290), 
Saint-Germain-de-Tallevende-la-Lande-Vaumont (F-14500), and 
Beaujeu-Saint-Vallier-Pierrejux-et-Quitteur (F-70100). I became curious about their specificity as intercessors to God,  and their  popularity.

This led me to (numerical) questions: the first being,  how many cities, so called hagiotoponyms,  in France bear the name of a saint.  Is there a preferential name distribution?, for example as  discussed in \cite{kulaSimmel} for babies names.
 
I decided to download the list of all cities in France, also thereafter sometimes called "communes".

   \begin{figure} 
    \centering
\includegraphics[height=13cm,width=13cm]{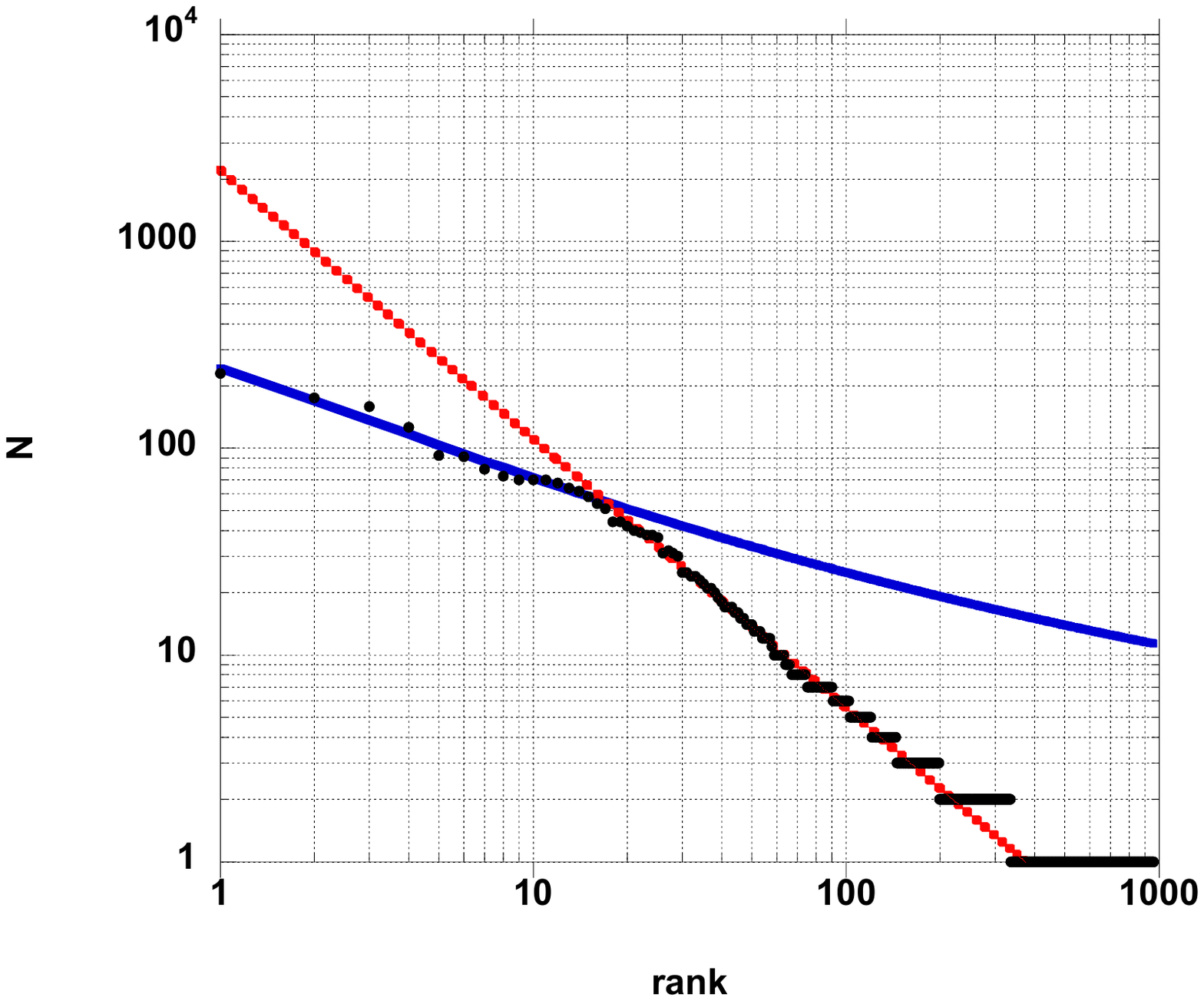}
\begin{centering}
 \caption{Log-log plot of the number  ($N$) of  hagiotoponyms in France  starting with  either Saint-  or Saint,  and referring to a supposed or confirmed  catholic saint  from INSEE data bank;  the best power law fits are shown when separating high and low rank cases.}
 \end{centering}
\end{figure}\label{FZipf3}
\newpage

\newpage
{\it In memory of Dietrich Stauffer}
\section{Introduction}
Statistical mechanics literature is full of reports, discussions, measurements, about
phase transitions in many varied systems, usually occurring between an ordered and a disordered phase.
Most of the relevant  phenomena occur as a function of the temperature \cite{Stanley}, thus in thermodynamics. External fields, like pressure or magnetic field, can influence the phase transitions pattern(s),  Other cases pertain to geometrical concepts or descriptions, like for "percolation phase transitions" \cite{Stauffer}.

For completeness, let it be recalled that one can distinguish static phase transitions (drastic, even discontinuous, changes in the stationary profiles of the system through some "order parameter")  \cite{Stanley}, and   dynamical phase transitions (discontinuous changes in the  characteristic "relaxation times" of the system) \cite{roshani2004static,riego2018towards}. One has also   observed instabilities with inhomogeneous stationary states, needing  extension of the theory of thermal fluctuations around equilibrium states to non-equilibrium situations  \cite{bedeaux1977ballast,ausloos1981continuously,ausloos1982electrothermal,riego2017metamagnetic}.

Nowadays, statistical mechanics has evolved from such major fields and is touching more "exotic subfields", like in econophysics and sociophysics, for which "agent based models" are relevant. One has found analogies with  thermodynamics phase transitions  in   topics outside physics,  as in financial crashes  \cite{AuslooscrashPhA,ausloos2002crashes,sornette2017whystock}   or  opinion formations \cite{OpinionsStauffer,OpinionsGalam}.

I have found, as demonstrated below, a (herding-like, as I will propose) system, with two quite distinct phases, as we usually observe in organized-disorganized societies, - similar to "ordering transitions" in liquid crystals \cite{Singh00}, or in magnetic systems \cite{Belovetal76}. The matter concerns hagiotoponyms, the names of cities bearing a saint name in France.

France is a rather catholic country, with famous pilgrimages,  in cities which do not necessarily carry a saint name. However, devotions to saints have led to many hagiotoponyms, since the 4th-5th centuries.   Cities like St.  Etienne or (Mont) St. Michel are rather well known, and the Saints also.  
 In contrast, St. Tropez, St. Nazaire, St. Lo, St. Malo,  St. Emilion occur more rarely,  are also well known, but not much the Saints. Other cases, e.g., St. Saturnin, - yet sometimes called St. Sernin, or San Savournin, are not rare but are not "common".   Moreover, the saints "having given" their names to cities, can be males, but  such saints can also be females: indeed, there are many well  known, like Ste. Marie cities: Saintes-Maries-de-la-Mer, Moustiers-Sainte-Marie, ... or {\it a contrario} not so well known, like Ste. Verge, Ste. Thorette,  or Ste. Néomaye.


There are  34970 "communes" (including 17 in  Mayotte, and 212 in "France d'Outre-Mer", DOM-TOM) on Jan. 01,   2019, according to one of the (most recent) official (INSEE)  lists of cities in France\footnote{$https://www.insee.fr/fr/statistiques/4277602?sommaire=4318291$; one can also get some help from $https://fr.wikipedia.org/wiki/Listes_-des_-communes_-de_-France$ and  from the Michelin Guide}.  I   selected those which start with "Saint",  see Sect. \ref{data}, as it seems to be  a good choice {\it a priori}.  Soon, I realized that I should use  "Saint". as a string,  instead, - since the string and reference to some saint can occur inside (Coise-Saint-Jean-Pied-Gauthier) or at the end (Fleurey-l\`es-Saint-Loup; Pont-Sainte-Maxence) of some city name, - or in a "complex way" (Saint-Martin-Lars-en-Sainte-Hermine, Saint-Maximin-la-Sainte-Baume). Some clean up must be made visually, since for example "Saintes" is not the plural of  "Sainte" (but refer to a tribe living  in the area a long time ago), - thus should not be  considered in the relevant list. Yet, looking for cities fully devoted to a saint or sainte, I felt the necessity of including those referring to  "Our Lady" (Notre-Dame). Also   "Saint-Sauveur" is included, since referring to "somebody", - who is found not to be necessarily Jesus-Christ, -  the latter   is not considered to be a Saint. 

I did not consider "Dieu", as pertinent, though many cities contain such a  string (La Chaise-Dieu, Dieulefi, Villedieu ...). I kept angels, e.g., St. Michel; other debatable cases are discussed in the Data gathering section (Sect. \ref{data}), and in a "technical" Appendix. However, it can be mentioned here       that I have also decided that archangels and angels are "saints"\footnote{even though they are said to have no sexual attribute}, and males. 

However, for completeness, due to the existence of dialects\footnote{e.g., occitan, breton, .. see Appendix on the latter} one has to take into account, cities with names starting with "San", like San Savournin, who is nobody else that Saint Saturnin, pointing to some further work in order to select the right Saint set at counting time. For hypercompleteness, one should notice that  in Corsica several    cities have a saint name:  in  "Haute Corse", (Northern Corsica), there are 20 cities with "San";  
but one can distinguish "San-" (9 cases) , "Santa-" (6 cases), and  one should be aware that if the first letter of the saint is a vowel, one should consider "Sant' " (3 cases) in the data gathering,  and "Santo-" (2 cases), but there is also  Sainte-Lucie-de-Tallano, thus 21 hagiotoponyms (out of  236 communes)

In contrast, 
in "Corse du Sud" (Southern Corsica),  out of 124  communes, one finds San-Gavino-di-Carbini, 
Sant'Andr\' ea-d'Orcino,
Santa-Maria-Figaniella,
Santa-Maria-Sich\'e, but there is also Saint-Florent.

 One should add that  "saint(e)" is sometimes replaced by "don" or "dan", or "dame", for example in Dampierre, Dommartin, Dammarti, Dammarie, Dannemarie, Dame-Marie. 
 
  Notice also that I do not wish to distinguish {\it bona fide} catholic saints, whom the Catholic Church has canonized as saints, as listed in some Catholic Encyclopedia or  in The Oxford dictionary of Saints   \cite{farmer1987oxford}, or elsewhere\footnote{  $https://en.wikipedia.org/wiki/List_-of_-saints$; ; $https://en.wikipedia.org/wiki/List_-of_-early_-Christian_-saints$; 
 
  $https://www.catholic.com/encyclopedia$} from so called saints who have been locally defined as such because of some  specific recognition and for whom  the recognition does not extend to the worldwide church. 
   
These  remarks  are made in order to indicate the complexity of sorting out such hagiotoponyms, and to suggest that one has also to dive into some relevant  "saint"   information\footnote{$https://en.wikipedia.org/wiki/List_-of_-Catholic_-saints$}. One should be aware that one should  study city names  before counting them when  a "saint" string occurs in the name of a city; another example: "Les  Saintes-Maries-de-la-Mer", which in fact also refers to several (female) saints.

Possible unintentional errors in counting, with their  correction, sometimes based on assumptions, are  discussed    in the Appendix "Methodology and Materials". The problems (and "solutions")  are so listed in a somewhat arbitrary order as sources of concern on the   reliability of the final data  which has been analyzed. Some arbitrary choice  sometimes had  to be  introduced.  Previous experience on identifying hagiotoponyms  when examining financial aspects and demography  in Italy has been of much help \cite{SSD15RCMAseanalysisIThagio,MARCcliomQQ}.  in fact,  much trial and error procedure   was part of the research work. There is no guarantee that all Saints are found nor distinguished. However,  I have  a strong  "belief" that the sampling error bar is rather weak and  does not drastically  impair  the qualitative  findings. 

Thereafter, one can use a classical (Zipf-like) rank-size plot, or more complicated (but better) empirical laws \cite{pone.0166011roymaunivlaw,JAQM} .

\section{Data}\label{data}
 Several steps were to be  taken  when I downloaded  the name of French cities; I  saved names containing the string:   "saint", "sainte","saint-", "sainte-", "san", "sant", "santo", "santa", automatically including those like   "saints",  "saintes",   "-saint-" ,  "-sainte-", "-saints-" and "-saintes-".  One has to distinguish city names containing or not a hyphen in their name or not.

 I crosschecked the list with respect to the list of communes in  each French Metropolitan department\footnote{
 $https://fr.wikipedia.org/wiki/Projet: Communes_-de_-France/Noms_-des_-articles_-de_-commune$, 
 
 for the first 45 departments; the  1997 list is more complete, and more useful,  in :  
 
 $ https://fr.wikipedia.org/wiki/Projet:Communes_-de_-France/Noms_-des_-articles_-de_-commune$
\#$Liste_-2019_-(mise_- $\`a$_-jour_-le_-9_-septembre_-2019)$ }.  N.B. There are 101 departments in France: 96 in "Metropole", counting 2 departments in Corsica, and 5 away from the "Metropole", in DOM-TOM. In order to remain within a rather coherent framework, I decided not to take into account in the analysis the hagiotoponym cities in the 5  DOM-TOM departments, - where there are several hagiotoponyms in fact\footnote{"for completeness",  there are  7 female (5 different ones) and  21 male (16 different ones) hagiotoponyms, excluding Saint-Esprit};  of course, Mayotte was also self-excluded, since it  is a muslim department, - without saints and with specifically different  culture and laws.
 
 I checked also the  "Mairie" ("City Hall") list. There are cities in France without inhabitants, hence without a mayor, but having a City Hall, even though it is abandoned, but  it exists legally. 
 
 I verified the "postal code" and the "INSEE communal code"; sometimes the postal code does not correspond to the department; this  demanded further crosschecking.
 
 In so doing, multiple equivalent entries were expected to be removed, and missing entries in some data bank were hopefully taken into account.
 

  A comment is in order concerning data banks which were (or are) used:
  \begin{itemize}
 \item     data bank 1:  $acp2$  from $http://www.phpmyadmin.net$ ; this is an absolute junk;  a waste of time; cities are missing; due to likely (stupid) encoding when two cities have the same name;
  \item     data bank 2: official INSEE,  giving the list of "city halls", but by alphabetical order at first; thus it seems easy to get a list of cities commencing by S and within such a ''box'', to get cities commencing by   Sainte-, then Sainte ; then Saint-, and finally Saint. One should possibility  count and select combinations like '' -Saint- ''; however it is obviously not  no easy   to get cities  like ''Xyz-Saint-Lmn..''; the search being too tedious;
   \item     data bank 3: Liste des communes de France, from a Wikipedia project  gives the number of cities per department, within  an alphabetical order; thus more easy to select places like  '' Xyz-Saint- ''; however it is observed that several cities are sometimes missing; but since the alphabetical order is combined with a list of INSEE codes which approximately follow the alphabetical order, it is possible to guess what the missing city is, thereafter going to search for the code on a complementary data bank of INSEE, like in the case of  "cities which have changed names", for example, or have fused.
   Indeed  France cities change names once in a while\footnote{
   see
    $ http://insee.fr/fr/methodes/nomenclatures/cog/$
   $ recherche_-historique.asp?debut=1930\&fin=2011$}, fused, defused,  change departments, etc.
   
   
    \item     data bank 4: Michelin Guide, of interest, but the cities listed in the index are not necessarily those having a mayor nor a city hall, but the guide index is   useful. 
    
   \end{itemize}
   Notice that the expected alphabetical order is  (surprisingly) different in such data banks,  in brief depending on abbreviations, e.g. Ste. B...., for Sainte B..., thereby appearing after
   Sainte-M..., due to alphabetical "ordering": a , b, ..., e, ...
   
Thereafter, came the treatment of several ambiguities, leading to several
municipalities exclusions. The criteria for exclusion have been basically
due to (i) names which do not "obviously" point to a specific  human Saint
and/or (ii) whether the toponym, though being a "sanctified location", is
clearly derived from some Bible fact or event: e.g., Sainte-Eglise, Sainte Foi, Sainte-Baume, Sainte-Croix, ...

A related question concerns the "least popular" Saints, e.g. those occurring only once, so called hapaxes. I find interesting to ask
how many of them exist?  but  "why are such Saints rare?" is outside the purpose of the present report. Nevertheless, how they
geographically distributed is an unexplored  and interesting question to be  examined in further work.

Thus, the  100\% identification of a catholically recognized Saint is admittedly sometimes not possible.
A Saint having given his/her name to a city is not necessarily a {\it bona fide} catholic Saint, but
only a "Saint by tradition" (see    \cite{farmer1987oxford}).
This  ambiguity seems not relevant to my consideration.

Therefore,  the Saint name in the studied  list  is assumed to point  to a unique  Saint who  is then considered to be  the representative element of the related set of the Saints with the equivalent name.
 
    \begin{figure} 
    \centering
\includegraphics[height=14cm,width=14cm]{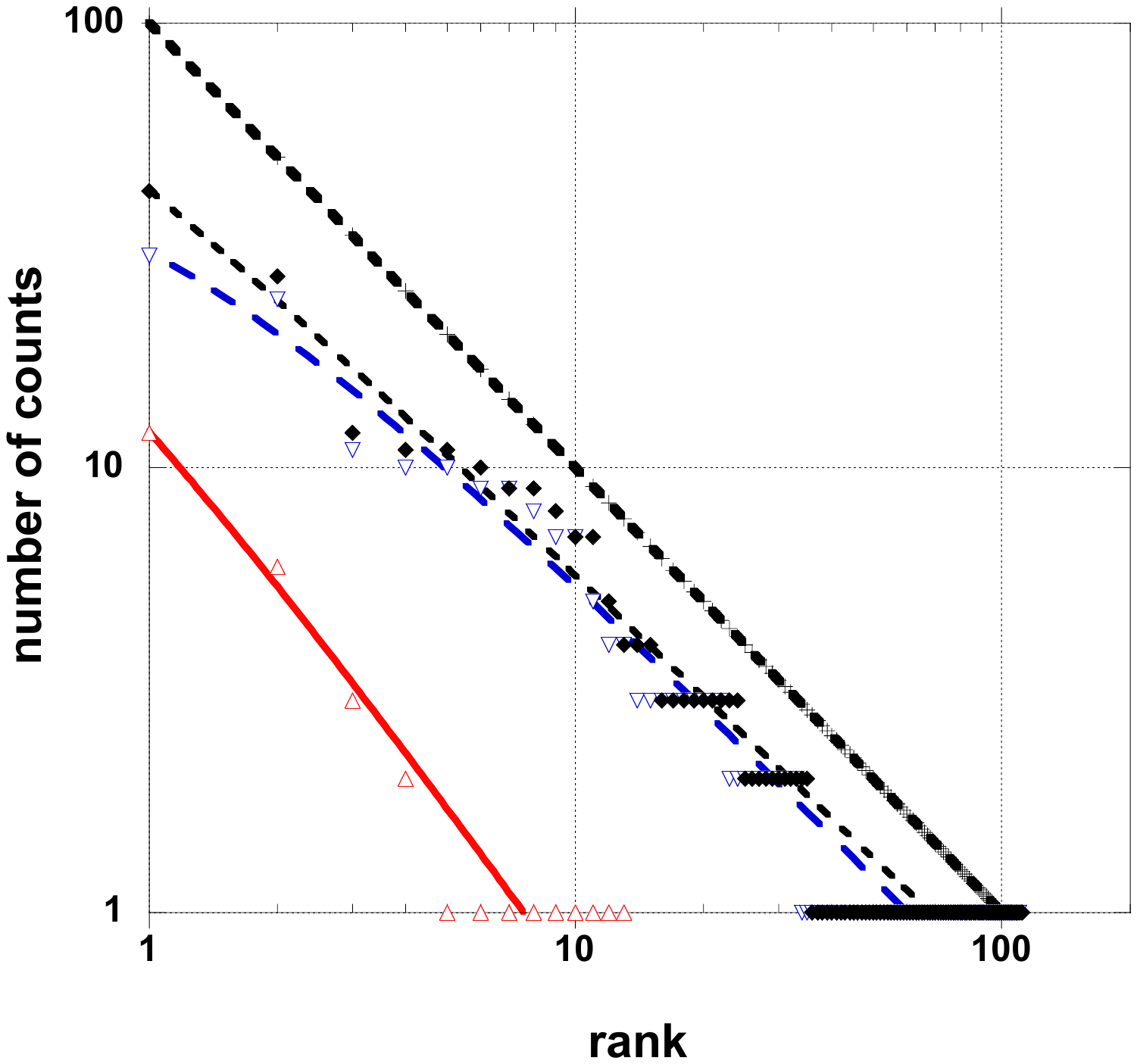}
 \caption{Rank-size log-log plot of the number  ($N_r$) of  hagiotoponyms in France   with  either  a male (blue,  down triangle) or a female (red, up triangle)  name,  and their combination (black diamond), derived from INSEE data bank; indicative empirical  fits are : based on a  BZM law (blue dash line)  or a mere power law (black tiny dot line) for males, and for females (red continuous line). Parameter values are given in the text. A  -1 power law  (black heavy dot line) is also shown for reference. }\label{Fallsaints}
\end{figure}

   \begin{figure} 
    \centering
\includegraphics[height=14cm,width=14cm]{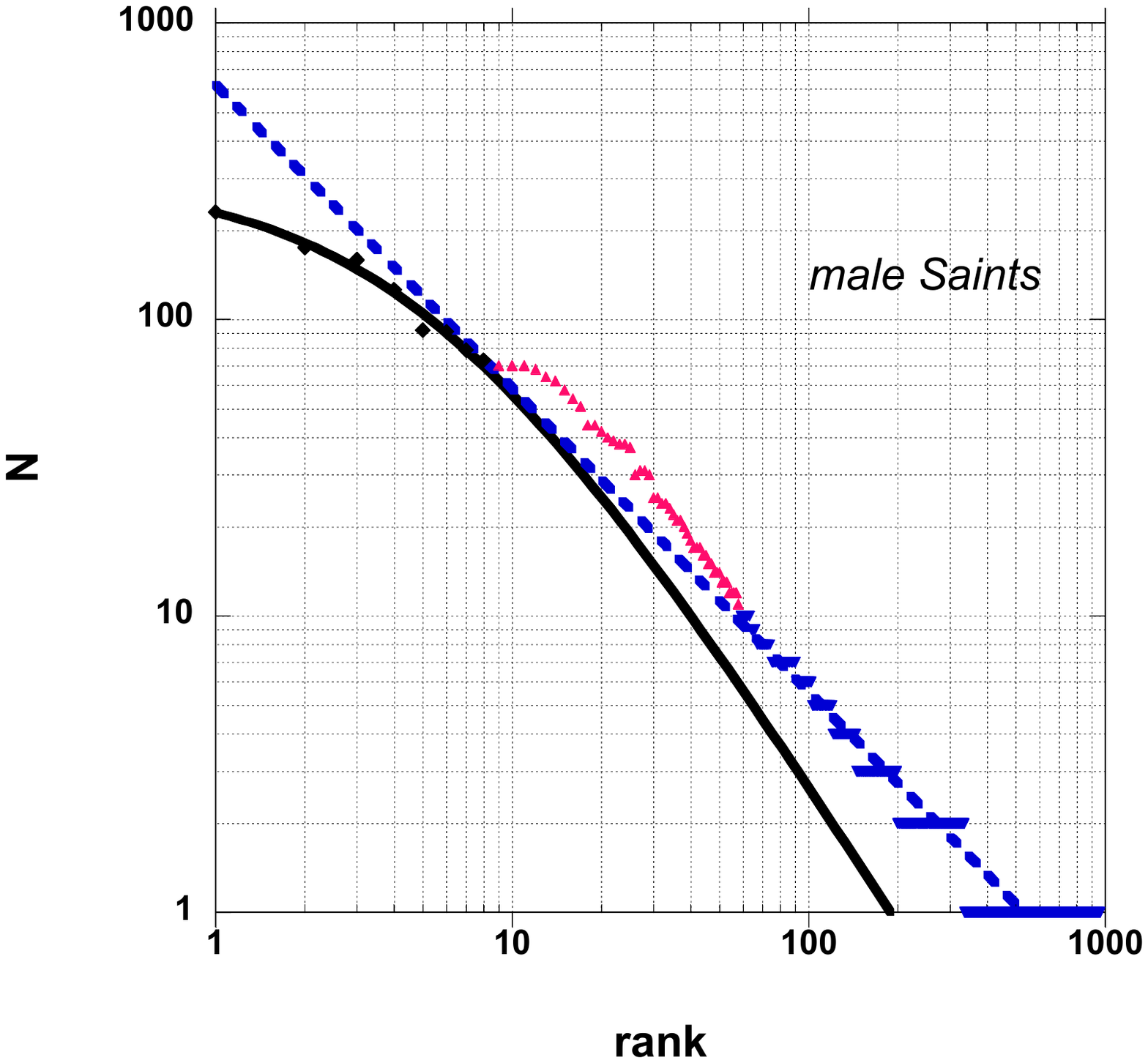}
 \caption{Rank-size log-log plot of the number  ($N_r$) of  male saints having their name included in that of cities in France    according to INSEE data bank; "best"  fits at low (a BZM  law; continuous line;  black dots) and high (a power law; dotted line; blue dots) ranks, respectively  are shown; an intermediary regime (red dots) is emphasized. Fit parameters are given in the text.}\label{Fhighlowmale}
\end{figure}

  \section{Data Analysis}

 The first  concern is the search for the statistical distribution of names of Saints attached to city names.    One ranks the saints by the number of towns using their names: $ N_1$ is the number of towns associated to the saint who has the most towns; $N_r$ decreases with $ r$; the alphabetical order is used if two saints have the same  $N_r$.  Thereafter,  a rank-size law is looked for.

Several sets of plots  can be made  available, with various axis types and scales, after counting the  cities having   various strings at various places in the hagiotoponyms.   %

Two cases illustrate the more interesting features. First, I present the log-log plot of the number  ($N_r$) of  hagiotoponyms in France, ranked in decreasing size order,    with  either  a male or a female  name,  and their combination (black diamond), derived from INSEE data bank, on Fig. \ref{Fallsaints}. The data is not far from a power law, called Zipf law \cite{z1}
\begin{equation}\label{Zipfeq}
N_r = \cfrac{N_1}{r^{\alpha}},
\end{equation}

 but is better represented by a Zipf-Mandelbrot-like (ZM) law, sometimes called Bradford-Zipf-Mandelbrot-like (BZM)  \cite{FAIRTHORNE}, 
 
   \begin{equation}\label{ZMlikeCr}
  N_r =\frac{J^{*}}{(\nu+r)^\zeta},
  \end{equation}
might be considered as more realistic.  It implies three parameters ($J^*$, $\nu$ and $\zeta$).

The fits are in each case made through a Levenberg-Marquardt Algorithm \cite{ Levenberg44,Marquardt63,LMalgorithm}. 
A mere power law for the female saint would be quite fair ($\alpha\simeq1.01$; $R^2\simeq 0.989$) and quasi  not distinguishable from a BZM law, Eq.(\ref{ZMlikeCr}),  with parameters $\zeta \simeq 1.02 \pm 0.06$; $R^2\simeq 0.992$), 
The overall fits  to a BZM law give   
  ($\zeta \simeq 0.978 \pm 0.04 ; R^2\simeq 0.984$), and   ($\zeta \simeq 0.918 \pm 0.032 ; R^2\simeq 0.989$),
  for the male saint and the overall combination, respectively.  For completeness, let it be mentioned that the "king" \cite{EPJB2.98.525stretchedexp_citysizesFR}  is St. Martin, and the "vice-roys" \cite{cerqueti2015evidence}, St. John and St. Peter.
  
  A  $r^{ -1}$  power law  is also shown  on Fig. \ref{Fallsaints}, reminiscent of a similar value found in a related problem with hagiotoponyms in Italy \cite{SSD15RCMAseanalysisIThagio,MARCcliomQQ}.  
  
A second case indicating the complexity of the resulting analysis is  a rank-size log-log plot of the number  ($N_r$) of  (only)  male saints having their name attached  to that of cities in France    according to INSEE data bank; the size is ranked in decreasing order on Fig. \ref{Fhighlowmale}. One can observe two smooth regions:   at low  rank, a distribution, well described by a   BZM law, and at high rank,  by a mere power law. In between the  behavior is more complex, as often in such plots, or systems. The main fit parameters are   equal to ($\zeta \simeq  1.579  \pm 0.091$; $\nu \simeq 5.21$;   $ R^2\simeq  0.991$),  for the low rank data, 8 data points, and   ($\alpha \simeq 1.0242 \pm 0.0007; R^2\simeq 0.979$), for the high rank data,  904 data points. It is fair to indicate that the  saints contributing to the low rank fit are St. Germain St. Julien, St. Laurent  St. Hilaire and St. Aubin. For completeness, let it be mentioned that after  the "king" (!) \cite{EPJB2.98.525stretchedexp_citysizesFR}, Ste. Marie, the  "vice-roy"  \cite{cerqueti2015evidence}, is Ste. Colombe. 

However, it occurs to astute readers and tourists that one should  distinguish hagiotoponyms not necessarily according to the linguistics aspects, but according to the "relevance" of the saint in the commune name. One can propose to consider that if the name of the city begins with the string "Saint", the  religious aspect of the localisation has a more relevant religious content, that if the commune name "ends" with  some  string "-Saint-Xyz...". In the latter case, the beginning of the name of the commune is the most relevant term, - the saint name, coming next is an additional information, in order to distinguish, e.g., previous parishes or localities.  As an example, consider Nuits-Saint-Georges:  "Nuits" is "more important" than St. Georges. Likewise, in St. Saturnin-lès-Apt and St. Saturnin-lès-Avignon, two nearby to each other  cities in Vaucluse department,  "-lès-Apt" and "-lès-Avignon" are mere indications of the area, - the emphasis being on St. Saturnin.

Therefore, the previous  method can be  reformulated  on such a  filtered sample, taking into account only cities "starting with a saint name". This leads to the (amazing) Fig. 1. The power law exponent is ($\alpha   \simeq 0.454 \pm 0.001 $; $R^2\simeq 0.983$) and ($\alpha \simeq  1.30\pm 10^{-10}$; $R^2\simeq  0.999$)
at low rank  and high rank respectively.
The critical rank is $r_c\simeq 16 \pm 1$.

Finally, a  numerical value on hapaxes is of interest: there are 628 "male hapaxes", for a total of 962 different male saints, and 77 "female hapaxes", for a total of 296 female saints. This points to a  very large variety of specific devotions and intercessors to God, often by  persons not recognized as duly sanctified along catholic rules. 

 \section{Agent based model sketch}
 
 The features of Fig. 1, power laws,  the critical rank value, and the ''huge'' number of hapaxes,  suggest some reasoning, whence  to sketch a socio-psychological  model.
 One has  a set of agents which (must) have an opinion, the name of  a preferred  saint; such agents  (devoted citizens or their political or religious leaders) prefer to be  
 with a saint, well known, who has a good reputation as being a fair intercessor to God; on the other hand, there are other agents who prefer to have a  more personal saint, less popular, but more to their living place, often a sort of local hermit, who was a healer or benefactor; those agents are prompt to indicate that in their  living place they have a specific  (say hermit-like) "holy" benefactor, also with good connection to God since he  is such a practical proof doing something good to them.  In fact, this might be a reason why visually (and practically "touristically")  one observes that the hapaxes saint are closely  distributed. One can consider that a rivalry exists between holy persons, saints by extension,  being links to God, - or by the local villagers.
 
This herd behavior is in line with the model first proposed by   Banerjee in 1992 \cite{Banerjee}. He proposed a herding model of decision making: in which people are inclined to mimic others’ actions, collectively, but sometimes at a price.   One may also follow a leader, whence  becoming a "follower" \cite{mosquera2015follow}. However, it is possible that one accepts others’ idea(s) even though one's own information tells  to do something else  \cite{guo2009herd,kononovicius2013threestates,dong2008self,kononovicius2014continuous,schweitzer2013canherding}.

More recently,   Eguiluz and  Zimmermann, 
 \cite{EZmodel} proposed an important application of herding model in information transmission with application to financial markets.
The $ N$ agents are on $N$ vertices of a network. Several agents   connect with each other to form a cluster for "some reason", but  sharing the  same information or opinion, in the present case about the interest of having a saint name for a location.  The degree heterogeneity of the network would affect the herding behavior and would lead to an order-(dis)order transition \cite{lambiotte2007howdoes}, as observed here. 
 
  Observe that one observes the endogeneous emergence of ''leaders'' in  populations, like for the name of babies, as already mentioned \cite{kulaSimmel}. In standard   opinion dynamics models, herding behavior is usually considered to be obeyed at some local scale, due to the interaction of
single agents with their neighbors. At more global scales, such models are governed by purely diffusive processes, as for example observed in universe creation discussion     \cite{FrontiersMA}.  In the present study,    some (somewhat external) agents influence  the others and  induce a  sort of phase transition in diffusive phases. This leads to  a herding phase where a fraction of the agents self-organizes into a regime  leading to a  rather global opinion, and to another less herding phase in which  the whole population prefers  another self-organization.  
  
 A more grounded historical study should take into account that the time scale is rather known, since most of the hagiotoponyms appeared in early christianization time (4th or 5th century) of the Western Europe. Of course, city names are changing throughout centuries in France, but as far as I could see,  do not much modify the numerical inputs in recent times (after Napoleon organization of France into Departments).

Should one say that the population opinion for or against a given saint, at the early times, could  lead to challenges and lead to different tails in the saint name opinion distribution \cite{Gurjeet2CSF}, according to the preference "{\it herding or not herding, that is  the question}". Notice that for being closer to reality,  one should generalize the  considerations of Crokidakis and others  \cite{crokidakis2017nonequilibrium,schwammle2007different,xie2002finite} on opinion dynamics to networks rather than remaining on regular lattices. 

Thus, the ad hoc modelization of the findings should result from a highly complex generalization of the Banerjee-Eguiluz-Zimmermann (BEZ) model. This is outside the present aim, and is left for future work.

\section{Conclusions}\label{conclusions} 

On one hand, a phase transition, in thermodynamics, admittedly occurs, or is usually defined, when some ''order parameter'', usually a local variable,  shows some drastic change; often it vanishes. At a usual second order phase transition one finds similar ''critical exponents'' on both sides of a ''critical  point'' for  quantities measuring correlations of the system variables, because of ''fluctuatins''.  Usually such a ''critical point'' is a specific value of some continuous variable, like the temperature. Nevertheless,  one may have a succession of ''critical points''. in between these, various regimes may occur. They maybe characterized by power laws \cite{PRB66.02.174436LHubert}.
The same is seen in percolation studies \cite{Stauffer}.

On the other hand,
 the practice of venerating holy
figures (and their relics) is  a cultural phenomenon that engaged all sections
of society \cite{MorrisTerritory}. Saints enjoyed high levels of popularity through their cults, leading to defining cities by the name of the saint whose cult is of interest.
Not all instances were the same.   
 Each city may have several parishes; each church is dedicated to a saint; communes bear the name of one or several saints.  There is such an important tradition in catholic  France. This leads to a very complex survey. In this research, one has been 
   looking for cities,  "fully devoted" to a ''Saint'' or a ''Sainte'', - not parishes.  
   
     The most  popular Saints  set is made among 16 individuals or so. In contrast, there is a huge set of Saints which are hapaxes, i.e. their name occurs only once.
 
 In the present work, one reports measures of the intensity (''size'') of an ''order parameter'', i.e. the number of times a ''saint popularity'' has led to a hagiotoponym in France. One looks for the intensity decay as a function of its  ''probability of occurrence'', personified by its rank in the list. One obtains some straight line on a log-log plot, with values of power law exponents ($\zeta$), different on both sides of a ''critical rank'' ($r_c$), i.e.  $\zeta \simeq 0.45 $ and $\zeta \simeq 1.30$, at low and high rank respectively, with $r_c =16$.
 
 I do not claim that one should distinguish between a mean field and a critical regime, where $r_c$ would rather be corresponding to the Landau point separating the mean field and critical regimes. The rank values are discrete points; this  leads to a distinctive difference with thermodynamical phase transitions. However, the rank distribution of such a saint importance follows simple power laws, as those found in many phase transitions aspects. This has allowed some reasoning in order to propose that the features are those of  a social phenomenon which  can be called a herding phase transition.

One could somewhat validate such a claim  showing clustering for at least certain names, i.e. the most popular ones.   For the saints in the core, one could be extending  the notion of popularity à la Hirsch \cite{Hirsch,Auslooscore}) through a core index, the h-index, being $r_c$. Below the critical rank, one could do a Voronoi  tessellation    \cite{TASM} for each main saint, - obtaining the average distance between the nearest neighbor hagiotoponyms, for further conviction. This can also be left for future work.

In conclusion, one should not be afraid of limitations, and assumptions in pursuing this work. Nevertheless, there are interesting open questions. For example, what kind of network is the embedding set? What is the distribution through departments. Can this be related to geographical, sociological, political, historical facts? And {\it in fine}, it seems  a hard  job but an interesting challenge to look at parishes and church dedications, - and their life time.
 


\newpage
\section{Appendix: Methodology and Materials}\label{method}
In this Appendix, I list   "problems" and solutions I found when  selecting the sample for further analysis.

For example, it occurred to me at once that one should be  aware that  words containing "saint" were not referring to some individual. The most obvious (to me)  was  the "Saintes" city (see below).

Moreover, 
 \begin{itemize} 
    \item  "Saintes"  is not referring to  several female saints, but to the local tribe  2000 years ago in the area; so Saintes has not been counted; idem for "Saints" (-en-Puisaye) (89520)
 \item idem Saints  F-77120 and F-89520 have not been counted
 \item  but Saints-Geosmes makes life complicated, because it is only one city, but referring in one word to three (twin) saints; Geosmes being an alteration of $jumeaux$; this  exceptional case (city) has been kept as referring to only one saint; not a drastic point
 \item  the mention of some  "saint(e)" is sometimes replaced by "don" or "dan", or "dame", for example in Dampierre, Dommartin, Dammarti, Dammarie, Dannemarie, Dame-Marie; but that increases much the number of possible strings, and demand to research many cities; I have neglected such strings
 
\item I have neglected the breton toponyms, containing $lok$, but not meaning "saint'' even though sometimes referring to a name like that  of a saint:  Locmaria (4 cases), Locronan surely referring to St. Ronan), Loctudy (likely referring to St. Tudy); notice that St. Tudy is  himself sometimes identified with St. Tugdual (1 case), St.  Pabu (1 case), St. Pabut (0 case), or St. Paban (0 case); St. Tugdual  and  St.  Pabu  were conserved;
  
   \item  cases like St. Saturnin = St. Savournin = St. Sernin,  have to be identified as only one  individual
   
    \item   one sometimes has to group hagiotoponyms; Saint Calais (F-72120) with Marolles-lès-Saint-Calais, also  F-72120;  notice two different city names, thus cities,  in France list but same postal code; there are other same saints most likely Saint-Calais-du-Désert (F-53140);  
    
    \item one has sometimes to "ungroup" hagiotoponyms: Saint-Remy-en-Bouzemont-Saint-Genest-et-Isson (F-51290) into two different saints; N.B. the longest name of a city in France;
    
     \item one  can encounter a full name with some other information : Nuits-Saint-Georges, Moustiers-Sainte-Marie, 
  Granges-Sainte-Marie; thus the string "saint" must be allowed to appear in various positions in a hagiotoponym
  
  \item too bad also for  Saintry-sur-Seine F-91250,  which is not related to Saint Ry; idem for Sainteny F-50500, 
    \item
    but, Saint-Genis-de-Saintonge  F-17240 should not be counted twice; Saint Onge does not exist! (see $https://fr.wikipedia.org/wiki/Saint-Onge$, for an illuminating  comment)


\item  a   problem  occurs without specific decision  for case like 
  Mas-Saintes-Puelles and Les-Saintes-Maries de la Mer ? ... how many saintes have to be taken into account?
   \item  "La  Chapelle-aux-Saints ' F-19120  and  "Longpré-les-Corps-Saints", F-80510 are   also removed   because they do  not pertain to   specific saints

  \item nor Nieul-lès-Saintes (F-17810); in french, "lès" means  "next", here "next to the city" Saintes;
  \item nor Mas-Saintes-Puelles (F-11400); they buried Saint Sernin (St. Saturnin), but their name and number is unknown!;

\item  finally, "interestingly", with respect to the model idea, in counting communes, several contain "saint" twice (8 cases);  recall the  ''spectacular'' case of Saint-Remy-en-Bouzemont-Saint-Genest-et-Isson (F-51290) mentioned here above; 
but, in addition, this shows that  the number of cities with a saint name is related but not identical to the total number of  used names.
\end{itemize}

 Other    warnings seem also  relevant!
   
 \begin{itemize} 
 
 \item (I) Eloy and Eloi are considered to be two different saints, but (II)  Andre and Andr\'e are the same saint, though there might be different Andre and Andr\'e, and (III) local names   have been used, identified or not, depending on my knowledge of the saint, i.e.   Savournin and Saturnin; 
that does  not seem to be a very drastic approximation though;
 

\end{itemize}
Improving on such three approximations would request to search about the saint itself, dive into his/her life, if any, and compare informations much outside those presently looked for. 
 
\vskip1.5cm
{\bf  NOTE ADDED  AFTER  ACCEPTANCE,  - added on galley proofs at production time:}
  
 Interesting support literature, in particular pertinent references on  hagiotopynms in France,  can be found in Ch. Higournet's work \cite{Hig1,Hig2}, 
  and in subsequent  reports \cite{Billy,Taverdet}.

 These studies are also to be found on the Internet site $http://mnytud.arts.unideb.hu/onomural/$




 

 \vskip0.5cm
{\bf  Acknowledgements}
 \vskip0.5cm
 Thanks to  A. P{\c{e}}kalski and K.  Ku{\l}akowski, for various comments; special thanks go to C. Berman and D. Berman.

 This research would not have been taking place if Dietrich Stauffer had not shown me his interest in exotic applications of statistical physics, and approved my working on sidelines. This paper would subsequently never have  been written, - nor published! Thanks again Dietrich for  illuminating scientific frontiers roads, with humour.  See you soon.

\end{document}